\documentclass[11pt,aps,prd,preprint]{revtex4}

\usepackage[dvips]{color}
\usepackage{epsfig}
\usepackage{amsmath}
\usepackage{graphicx}
\usepackage{subfigure}
\date{\today}

\usepackage[dvips]{color}
\usepackage{epsfig}
\usepackage{amsmath}
\usepackage{graphicx}
\usepackage{subfigure}
\date{\today}
\begin{document}
\title{  Kiselev/CFT correspondence and black hole thermodynamics}
\author{J. Sadeghi $^{1,5}$}
\email{pouriya@ipm.ir}
\author{B. Pourhassan $^{2,5}$}
\email{b.pourhassan@du.ac.ir}
\author{M. Rostami $^{3}$}
\email{m.rostami@iau-tnb.ac.ir}
\author{Z. Nekouee $^{4}$}
\email{z.nekouee@stu.umz.ac.ir}

\affiliation{ $^{1} $ \small{Department of Physics, Faculty of Basic Sciences, University of Mazandaran, P. O. Box 47416-95447, Babolsar, Iran}}
\affiliation{ $^{2} $ \small{School of Physics, Damghan University, Damghan, 3671641167, Iran}}
\affiliation{ $^{3} $ \small{Department of Physics, North Tehran Branch, Islamic Azad University, Tehran, Iran}}
\affiliation{ $^{4} $ \small{Department of Mathematics, Faculty of Basic Sciences, University of Mazandaran, P. O. Box 47416-95447, Babolsar, Iran}}
\affiliation{ $^{5} $ \small{Canadian Quantum Research Center, 204-3002 32 Ave Vernon, BC V1T 2L7 Canada}}
\begin{abstract}
In this paper, we study the thermodynamic properties of Kiselev black hole and its holographic dual.
We obtain the thermodynamic product formula for the Kiselev black hole. We consider Kiselev Black
hole surrounded by both dust and radiation. We find that the area (or entropy) product formula for both cases is mass-independent as well as the case in the Einstein gravity, which interpreted as universal quantity. Moreover, we calculate the black hole entropy bound for both the inner and outer horizons. Furthermore, we show that the central charges of the left and right moving sectors are not the same via universal thermodynamic relations. Such universal relations lead us to calculate the central charge of conformal field theory (CFT). Finally, we see that the left and right of central charge of CFT are same.
\end{abstract}
\maketitle
\newpage
\section{Introduction}
As we know, our universe is currently in accelerating expansion stage, in this situation dark energy is the best candidate of such acceleration \cite{1,2,3,4}. The dark energy produces strong repulsive gravitational effect and violets the null and weak energy conditions \cite{5,6}. Now, finding nature of dark energy is one of the important problems in theoretical physics and astronomy. In that case there are several models to describe dark energy \cite{D1,D2,D3,D4,D5,D6,D7,D8}. In that case, black holes are also interesting things in the context of accelerating expansion of the universe. General classification of black holes are two kinds of charged and uncharged cases. Each of them are of two types which are with and without rotation. Hence, four general kind of black holes are as Schwarzschild \cite{BH1} (uncharged and non-rotating), Kerr \cite{BH2} (uncharged and rotating), Reissner-Nordstr\"{o}m \cite{BH3} (charged and non-rotating) and Kerr-Newman \cite{BH4} (charged and rotating) black holes. For example, thermodynamics and statistics of Kerr-Godel black hole given by the Ref. \cite{pourdarvish} while thermodynamics of a sufficient small singly spinning Kerr-AdS black hole investigated by the Ref. \cite{Kerr-AdS}. One of the best way to obtain information about black hole is thermodynamics study. In that case there are several works where the black hole thermodynamics has been studied \cite{S1,S2,S3,S4,S03}, some of them used the geometric method to study the black hole thermodynamics. Ref. \cite{S2} is indeed extension of Ref. \cite{S5} to the case of rotating black hole. We know that the Bekenstein-Hawking entropy of any black hole is related to the outer event horizon area. In that case, microscopic origin of the black hole entropy obtained using the holographic principles \cite{JHAP1, JHAP2}. For example, the macroscopic entropy of black holes could be obtained from the microscopic states which is well-known as Kerr/CFT correspondence \cite{1103.2355}. In that case, one of the interesting problems is that the central charges of the dual picture of the Kerr black hole are mass-independent, which means that horizons area product of these
black holes are also independent of mass. It is confirmed for the four-dimensional Kerr-Newman black hole \cite{0810.3998, 0903.5405}. Similar result obtained for the general five- and four-dimensional multi-charged rotating black holes \cite{C1, C2, C3, C4, C5}. It has been also investigated for the higher dimensional black holes \cite{1011.0008} as well as three dimensional black holes \cite{Detournay} and suggested universal behavior. It is also interesting to consider two dimensional black holes \cite{2d1,2d2} and investigate the area product. Now, in this paper, among various kinds of black holes we would like to consider Kiselev black hole which is surrounded by various types of matter. In that case accretion onto a Kiselev black hole investigated by the Ref. \cite{K1}. Also, strong gravitational lensing of Kiselev black hole has been studied by the Ref. \cite{Azreg, Younas}. Moreover, thermodynamics
and phase transition of Kiselev black hole investigated by the Ref. \cite{majeed}  and products relating the surface gravities, surface temperatures, and Komar energies calculated. Recently, Tidal forces in Kiselev black hole examined by the Ref. \cite{Shahzad}. These are our motivation to study area product of Kiselev black hole surrounded by both dust and radiation and investigate universal behavior explained above. The thermodynaics of similar solutions already studied by \cite{N1,N2,N3}.
Indeed, similar to the Ref. \cite{20} we would like to study extremal Kiselev black hole by using Cardy formula and investigate the macroscopic Bekenstein-Hawking entropy formula. Ref. \cite{20} produces a thermal state with a temperature $T_{L} = \frac{1}{2\pi}$ for the Kerr black hole. We also derive some thermodynamic products and calculate the Smarr mass
formula for Kiselev black holes. Furthermore, we derive the area (or entropy) bound for both the horizons. Finally, by using the universal
thermodynamic relations, we show that the
central charges of the left and right moving sectors of the dual CFT \cite{200}
in AdS/CFT correspondence are not the same. Indeed, we tried to calculate
the central charge from the CFT for non-rotating Kiselev black hole while for
rotating Bardeen black hole have been already computed in \cite{N1, N4}.\\
So, in the next section we review some important area product formula. In the section \ref{III} we review special cases of Kiselev black holes. The first case is called radiation black hole. So, area product formula in Kiselev radiation black hole also considered in this section. Then, we study holographic dual for the Kiselev radiation black hole. In section \ref{III} also we repeat calculations for the Kiselev dust black hole. Finally in section \ref{IV} we give conclusion.
\section{Black hole thermodynamics}
Generally, one can say that the Bekenstein-Hawking entropy for the black hole is proportional horizon area which is given by,
\begin{equation}\label{s1}
S_{\pm}=\frac{A_{\pm}}{4},
\end{equation}
where units used in which $G=\hbar = c = \kappa = 1$ and $A_{\pm}$ are area of both
the horizons. Similarly, the Hawking temperatures are proportional to the
surface gravity of event horizon and Cauchy horizon. They are defined
as,
\begin{equation}\label{s2}
T_{\pm} = \frac{\kappa_{\pm}}{2 \pi},
\end{equation}
where $\kappa_{\pm}$ is surface gravity \cite{1, 2}.\\
In that case the first law of black hole thermodynamics for both horizons becomes
\begin{equation}\label{s3-0}
dM = \pm \frac{\kappa_{\pm}}{8 \pi} dA_{\pm} + \Omega_{\pm} dJ+\Phi_{\pm} dQ,
\end{equation}
where $\Omega_{\pm} = \frac{\partial M}{\partial J} $ is angular momentum, and $\Phi_{\pm} = \frac{\partial M}{\partial Q} $ is the electric potential.\\
Recently, several papers have shown that the area product formula for some black hole is independent of mass parameter \cite{3}. Also, Ref. \cite{4} for the black hole in $D=4$ and $D \geq 4$ obtained some new results which are satisfied by the quantization formula. It means that the area product (or entropy product) formula will be as,
\begin{equation}\label{s3}
A_{+}A_{-}=({8 \pi}\ell_{p\ell}^{2})^{2}N ,
\end{equation}
where $ \ell$ is Planck length \cite{51, 61, 71, 81, 91}.\\
On the other hand, the holography connect quantum gravity to quantum field theory \cite{10, 11}. As the fist example, we see the Kerr black hole is dual to CFT for the $AdS_{3}$ background \cite{12, 13}. They took some boundary conditions and calculated the important quantity in CFT as a central charge. So, generally they proved Kerr black hole dual to chiral CFT and the thermal state is with a temperature  $T_{L}=\frac{1}{2\pi}$ \cite{14}. So, we use the above information and apply to Kiselev black hole. We try to employ the Refs. \cite{C2, 15} and study the universal relations. In that case, the microscopic degrees of freedom of the black hole are illustrated in terms of conformal field theory. So, the inner $A_{-}$ and outer $A_{+}$ Killing horizon of black holes correspond to $N_{R}$ and $N_{L}$ which is given by,
\begin{equation}\label{s3}
\frac{A_{+}A_{-}}{({8\pi})^{2}}=N_{R}-N_{L},
\end{equation}
where $N_{R}$ and $ N_{L}$ are the number of right and left moving excitations of the CFT respectively.\\
There are several evidence that outer $A_{+}$ and inner $A_{-}$ are satisfied by following relations \cite{C4, 17},
\begin{equation}
\frac{A_{+}}{4} = 2\pi \sqrt{N_{R}} + 2\pi \sqrt{N_{L}},
\end{equation}
and
\begin{equation}
\frac{A_{-}}{4}=2\pi\sqrt{N_{R}}-2\pi \sqrt{N_{L}}.
\end{equation}
Also, the relation between right entropy and left entropy is given by,
\begin{equation}
\frac{A_{\pm}}{4}=S_{R}\pm S_{L}.
\end{equation}
So, here we want to investigate the thermodynamic properties of inner and outer horizon of Kiselev black hole for the radiation and dust cases \cite{19, 22}.
We explicitly verify the first law of black hole mechanics which is completely in agreement with the inner horizon as well as outer horizon. Moreover, we derive the Smarr mass formula for these class of black hole. Furthermore, we derive the area (or entropy) bound for both the horizons. Finally, by using the universal
thermodynamic relations we show that the central charges of the left and right moving sectors of CFT are the same \cite{23}. It should be noted the area product is mass-independent in Kiselev black hole. In that case the universal relation will be as $T_{+}S_{+}=-T_{-}S_{-}$.
\section{Kiselev Black Hole and Universal Relation}\label{III}
In order to discuss the holography, we have to consider the line element of state charge black hole. So, we introduce such element which is surrounded by energy-matter \cite{24}. So, we have following line element,
\begin{equation}
ds^{2}=f(r)dt^{2}-\frac{dr^{2}}{f(r)}-r^{2}(d\theta^{2}+\sin\theta d\phi^{2}),
\end{equation}
with
\begin{equation}
f(r)=1-\frac{2M}{r}+\frac{q^{2}}{r^{2}}-\frac{\sigma}{r^{3\omega+1}},
\end{equation}
where $M$ and $q$ are the mass and electric charge, $\sigma$ and $\omega$ are normalization parameter and
matter equation of state parameter around the black hole, respectively.  For this black hole, we consider two cases, first case $\omega=\frac{1}{3}$, in that case the corresponding black hole surrounded by radiation. In second case $\omega=0$, the corresponding black hole surrounded by dust.
\subsection{Radiation}
In that case we have $\omega=\frac{1}{3}$. For Kiselev black hole surrounded by radial radiation we have,
\begin{equation}
f(r)=1-\frac{2M}{r}+\frac{q^{2}}{r^{2}}-\frac{\sigma_{r}}{r^{2}},
\end{equation}
where $\sigma_{r}$ is parameter of radiation. The black hole horizon will be as,
\begin{equation}\label{12}
r_{\pm}=M\pm\sqrt{M^{2}-q^{2}+\sigma_{r}},
\end{equation}
where $r_{+}$ corresponds to event horizon and $r_{-}$ corresponds to Cauchy horizon. We will assume only the cases in which $M^{2}-q^{2}+\sigma_{r}\geq 0$. Because naked singularities $M^{2}-q^{2}+\sigma_{r}< 0$ do not occur in nature if the cosmic conjecture
is true. By using the equation (\ref{12}), product of their event horizon will be as,
\begin{equation}
r_{+}r_{-}=q^{2}-\sigma_{r},
\end{equation}
which is independent of mass and the area of this black hole is given by,
\begin{eqnarray}\label{14}
A_{\pm}&=&\int_{0}^{2\pi}\int_{0}^{\pi}\sqrt{g_{\theta\theta}g_{\phi\phi}}d\theta d\phi=4\pi r_{\pm}^{2}\nonumber\\
&=&4\pi(2Mr_{\pm}-q^{2}+\sigma_{r}),
\end{eqnarray}
and their product will be as,
\begin{equation}
A_{+}A_{-}=16\pi^{2}(q^{2}-\sigma_{r})^{2}.
\end{equation}
It implies that area product formula is independent of black hole mass. Thus area product formula in Kiselev radiation black hole is universal. By using the equation (\ref{14}), the mass could be expressed in terms of area of both horizons,
\begin{equation}\label{16}
M^{2}=\frac{\pi(q^{2}-\sigma_{r})^{2}}{A_{\pm}}+\frac{A_{\pm}}{16\pi}+\frac{q^{2}-\sigma_{r}}{2}.
\end{equation}
The Bekenstein-Hawking entropy is,
\begin{equation}
S_{\pm}=\frac{A_{\pm}}{4}=\pi(2Mr_{\pm}-q^{2}+\sigma_{r}),
\end{equation}
thus the entropy product formula is given by,
\begin{equation}
S_{+}S_{-}=\pi^{2}(q^{2}-\sigma_{r})^{2},
\end{equation}
it is also independent of mass. Thus the entropy product formula also universal in Kiselev black hole. In that case, the Hawking temperature is determined by using the following formula,
\begin{equation}
T_{\pm}=\frac{\kappa_{\pm}}{2\pi}=\frac{1}{4\pi}\frac{df}{dr}|_{r=r_{\pm}}=\frac{r_{\pm}-M}{2\pi r_{\pm}^{2}},
\end{equation}
and their product yields,
\begin{equation}\label{19}
T_{+}T_{-}=\frac{q^{2}-\sigma_{r}-M^{2}}{4\pi^{2}(q^{2}-\sigma_{r})^{2}}.
\end{equation}
By using the equation (\ref{19}), the surface gravity is given by,
\begin{equation}
\kappa_{\pm}=\frac{r_{\pm}-M}{r_{\pm}^{2}},
\end{equation}
and their product yields,
\begin{equation}
\kappa_{+}\kappa_{-}=\frac{q^{2}-\sigma_{r}-M^{2}}{(q^{2}-\sigma_{r})^{2}}.
\end{equation}
It should be noted that surface gravity and surface temperature product both depend on mass thus they are not universal. The Komar energy of the black hole is defied as,
\begin{equation}
E_{\pm}=2S_{\pm}T_{\pm}=r_{\pm}-M,
\end{equation}
and their product is,
\begin{equation}
E_{+}E_{-}=q^{2}-\sigma_{r}-M^{2}.
\end{equation}
Here, we see the energy product is not an universal quantity. The irreducible mass $M_{irr}$ of a non-spinning black hole which is related to the surface area $A_{\pm}$ which is given by,
\begin{equation}
M_{irr,\pm}^{2}=\frac{A_{\pm}}{16\pi}=\frac{S_{\pm}}{4\pi},
\end{equation}
the product of the irreducible mass at the horizons are,
\begin{equation}
M_{irr,+}M_{irr,-}=\frac{q^{2}-\sigma_{r}}{4},
\end{equation}
the above equation is also an universal quantity. Now we write the equation (\ref{16}) in terms of irreducible mass,
\begin{equation}\label{27}
M=M_{irr,\pm}+\frac{q^{2}-\sigma_{r}}{4M_{irr,\pm}}.
\end{equation}
 The specific heat for Kiselev radiation black hole is given by,
\begin{equation}\label{28}
C_{\pm}=\frac{\partial M}{\partial T_{\pm}}.
\end{equation}
In order to obtain the equation (\ref{27}) we need to calculate the partial derivatives of mass $M$ with respect to $T_{\pm}$. In that case, one rearrange the the equation (\ref{28}) as
$\frac{\partial M}{\partial T_{\pm}}=\frac{\partial M}{\partial r_{\pm}}\frac{\partial r_{\pm}}{\partial T_{\pm}}$. So, one can write following expression,
\begin{equation}\label{29}
\frac{\partial M}{\partial r_{\pm}}=\frac{r_{\pm}^{2}-q^{2}+\sigma_{r}}{2r_{\pm}^{2}},
\end{equation}
and
\begin{equation}\label{30}
\frac{\partial T_{\pm}}{\partial r_{\pm}}=\frac{2M-r_{\pm}}{2\pi r_{\pm}^{3}}.
\end{equation}
By using Eqs. (\ref{28})-(\ref{30}) we have the following equation,
\begin{equation}
C_{\pm}=\frac{2\pi r_{\pm}^{2}(r_{\pm}-M)}{2M-r_{\pm}},
\end{equation}
and their product is given by
\begin{equation}
C_{+}C_{-}=4\pi^{2}(q^{2}-\sigma_{r})(q^{2}-\sigma_{r}-M^{2}).
\end{equation}
Finally, the Gibbs free energy is given by,
\begin{equation}
G_{\pm} = M - T_{\pm} s_{\pm} = \frac{3 M - r_{\pm}}{2}.
\end{equation}
The universal phenomena give us motivation to investigate the holography in K/CFT. As we know the central charge play important role in holography side. So, we will drive the central charge $c_{R}$ and $c_{L}$ of the right and left moving sectors
of the dual CFT in K/CFT correspondence. We will prove that the central charges of the
right and left moving sectors same i.e. $c_{R}=c_{L}$ for K/CFT black hole. Also, we will determine
the dimensionless temperature of microscopic CFT from the above thermodynamic relations.
Furthermore, using Cardy formula, we will explicitly compute the right and left moving entropies for the Kiselev black hole in case of radiation as,
\begin{equation}\label{33}
S_{R,L}=\frac{1}{2}(S_{+}\pm S_{-}),
\end{equation}
as well as the temperatures in 2D CFT,
\begin{equation}\label{34}
T_{R,L}=\frac{T_{+}T_{-}}{T_{-}\pm T_{+}}.
\end{equation}
By using Eqs. (\ref{33}) and (\ref{34}), we have following equations,
\begin{eqnarray}\label{35}
S_{R}=\pi(2M^{2}-q^{2}+\sigma_{r}),\quad S_{R}=2\pi M\sqrt{M^{2}-q^{2}+\sigma_{r}},
\end{eqnarray}
and
\begin{eqnarray}\label{36}
T_{R}=\frac{1}{8\pi M}, \quad T_{L}=\frac{\sqrt{M^{2}-q^{2}+\sigma_{r}}}{4\pi(2M^{2}-q^{2}+\sigma_{r})}.
\end{eqnarray}
The left and right moving sectors of entropies could be written in form of Cardy formula as,
\begin{equation}\label{37}
S_{R,L}=\frac{\pi^{2}}{3}c_{R,L}T_{R,L},
\end{equation}
then the mass-independence of area product of the horizons implies the following equation,
\begin{equation}
c_{L}=c_{R}=24M(2M^{2}-q^{2}+\sigma_{r}).
\end{equation}
If we set $q=0$ and $\sigma_{r}=0$, together $4M^{2}=a$ where $a$ is rotational parameter, ther recover result of Ref. \cite{N1} with the zero quintessential intensity.
\subsection{Dust}
Now we back to case of $\omega=0$. For Kiselev black hole surrounded by dust, we have,
\begin{equation}\label{39}
f(r)=1-\frac{2M}{r}+\frac{q^{2}}{r^{2}}-\frac{\sigma_{d}}{r},
\end{equation}
where $\sigma_{d}$ is parameter of dust. The black hole horizon correspond to $f(r)=0$,
\begin{equation}\label{40}
r_{\pm}=\frac{2M+\sigma_{d}\pm\sqrt{(2M+\sigma_{d})^{2}-4q^{2}}}{2},
\end{equation}
where is $r_{+}$ corresponds to event horizon and $r_{-}$ corresponds to Cauchy horizon. We will assume only the cases in which $(2M+\sigma_{d})^{2}-4q^{2}\geq 0$. Because naked singularities $(2M+\sigma_{d})^{2}-4q^{2}< 0$ do not occur in nature if the cosmic conjecture
is true. By using Eq. (\ref{40}), their product yields,
\begin{equation}\label{41}
r_{+}r_{-}=q^{2},
\end{equation}
which is independent of mass. The area of this black hole is given by,
\begin{eqnarray}\label{42}
A_{\pm}&=&\int_{0}^{2\pi}\int_{0}^{\pi}\sqrt{g_{\theta\theta}g_{\phi\phi}}d\theta d\phi=4\pi r_{\pm}^{2}\nonumber\\
&=&4\pi((2M+\sigma_{d})r_{\pm}-q^{2}),
\end{eqnarray}
and their product will be as,
\begin{equation}
A_{+}A_{-}=16\pi^{2}q^{4}.
\end{equation}

The other relevant thermodynamic relations are,
\begin{eqnarray}
A_{+} + A_{-} &=& 4\pi [( 2 M + \sigma_{d})^{2} - 2 q^{2} ],\nonumber\\
A_{\pm} - A_{\mp} &=& 4 \pi [ ( 2 M + \sigma_{d})\sqrt{( 2 M + \sigma_{d})^{2}- 4q^{2} } ],\nonumber\\
\frac{1}{A_{+}} + \frac{1}{A_{-}} &=& \frac{1}{4 \pi q^{2}} \left[\frac{(2 M + \sigma_{d})^{2} - 2 q^{2} }{( 2 M + \sigma_{d})-( 2 M +\sigma_{d})^{2} + q^{2}}\right],\nonumber\\
\frac{1}{A_{+}} - \frac{1}{A_{-}} &=& \frac{1}{4 \pi q^{2}} \left[\frac{(
2 M + \sigma_{d}) (\sqrt{( 2 M + \sigma_{d})^{2} - 4 q^{2}}) }{( 2 M
+ \sigma_{d})-( 2 M + \sigma_{d})^{2} + q^{2}}\right].
\end{eqnarray}
The above thermodynamic relations of all the horizons may be used to
determine the Kiselev-dust black hole entropy in terms of Cardy formula
which provides some evidence for a AdS/CFT correspondence. It implies that the area product formula is independent of black hole mass. Thus the area product formula in Kiselev radiation black hole is universal. By using Eq. (\ref{42}), the mass could be expressed in terms of area of both horizons,
\begin{equation}\label{44}
M^{2}+M\sigma_{d}=\frac{\pi q^{4}}{A_{\pm}}+\frac{A_{\pm}}{16\pi}+\frac{2q^{2}-\sigma_{d}^{2}}{4}.
\end{equation}
The Bekenstein-Hawking entropy is,
\begin{equation}\label{45}
S_{\pm}=\frac{A_{\pm}}{4}=\pi((2M+\sigma_{d})r_{\pm}-q^{2}),
\end{equation}
thus the entropy product formula is given by,
\begin{equation}\label{46}
S_{+}S_{-}=\pi^{2}q^{4}.
\end{equation}
It is also independent of mass. Thus the entropy product formula also universal in Kiselev black hole. In that case the Hawking temperature is determined by using the following formula,
\begin{equation}\label{47}
T_{\pm}=\frac{\kappa_{\pm}}{2\pi}=\frac{1}{4\pi}\frac{df}{dr}|_{r=r_{\pm}}=\frac{2r_{\pm}-(2M+\sigma_{d})}{4\pi r_{\pm}^{2}},
\end{equation}
and their product yields,
\begin{equation}\label{48}
T_{+}T_{-}=\frac{4q^{2}-(2M+\sigma_{d})^{2}}{16\pi^{2}q^{4}}.
\end{equation}
By using Eq. (\ref{47}), the surface gravity is given by,
\begin{equation}\label{49}
\kappa_{\pm}=\frac{2r_{\pm}-(2M+\sigma_{d})}{2 r_{\pm}^{2}},
\end{equation}
and their product yields,
\begin{equation}\label{50}
\kappa_{+}\kappa_{-}=\frac{4q^{2}-(2M+\sigma_{d})^{2}}{4q^{4}}.
\end{equation}
It should be noted that surface gravity product and surface temperature product both depend on mass, thus they are not universal. The Komar energy of the black hole is defied as,
\begin{equation}\label{51}
E_{\pm}=2S_{\pm}T_{\pm}=\frac{2r_{\pm}-(2M+\sigma_{d})}{2},
\end{equation}
and their product is,
\begin{equation}\label{52}
E_{+}E_{-}=\frac{4q^{2}-(2M+\sigma_{d})^{2}}{4}.
\end{equation}
We see, the energy product is not an universal quantity. The irreducible mass $M_{irr}$ of a non-spinning black hole which is related to the surface area $A_{\pm}$ which is given by,
\begin{equation}\label{53}
M_{irr,\pm}^{2}=\frac{A_{\pm}}{16\pi}=\frac{S_{\pm}}{4\pi},
\end{equation}
the product of the irreducible mass at the horizons are,
\begin{equation}\label{54}
M_{irr,+}M_{irr,-}=\frac{q^{2}}{4},
\end{equation}
the above equation is also an universal quantity. Now we can write Eq. (\ref{44}) in terms of irreducible mass,
\begin{equation}\label{55}
M=M_{irr,\pm}+\frac{q^{2}}{4M_{irr,\pm}}-\frac{\sigma_{d}}{2}.
\end{equation}
The specific heat for Kiselev radiation black hole is given by,
\begin{equation}\label{56}
C_{\pm}=\frac{\partial M}{\partial T_{\pm}}.
\end{equation}
The same as Eq. (\ref{28}) we calculate the partial derivatives of mass $M$ and temperature $T_{\pm}$ with respect to $r_{\pm}$ which are given by,
\begin{equation}\label{57}
\frac{\partial M}{\partial r_{\pm}}=\frac{r_{\pm}^{2}-q^{2}}{2r_{\pm}^{2}},
\end{equation}
and
\begin{equation}\label{58}
\frac{\partial T_{\pm}}{\partial r_{\pm}}=\frac{(2M+\sigma_{d})-r_{\pm}}{2\pi r_{\pm}^{3}}.
\end{equation}
By using Eqs. (\ref{56})-(\ref{57}) we have the following equation,
\begin{equation}
C_{\pm}=\frac{\pi r_{\pm}(r_{\pm}^{2}-q^{2})}{(2M+\sigma_{d})-r_{\pm}},
\end{equation}
and their product is given by
\begin{equation}
C_{+}C_{-}=\pi^{2}q^{2}(4q^{2}-(2M+\sigma_{d})^{2}).
\end{equation}
Finally, the Gibbs free energy is given by,
\begin{equation}
G_{\pm} = M - T_{\pm} s_{\pm} = \frac{2 r_{\pm}- (2M + \sigma_{d}
)}{4}.
\end{equation}
Similar to the Eqs. (\ref{35}) and (\ref{36}) we have the following equations,
\begin{eqnarray}
S_{R}=\frac{\pi}{2}[(2M+\sigma_{d})^{2}-2q^{2}],\quad S_{L}=\frac{\pi}{2}(2M+\sigma_{d})\sqrt{(2M+\sigma_{d})^{2}-4q^{2}},
\end{eqnarray}
and
\begin{eqnarray}
T_{R}=\frac{q^{2}}{4\pi(2M+\sigma_{d})},\quad T_{L}=\frac{q^{2}[(2M+\sigma_{d})^{2}-4q^{2}]}{4\pi\sqrt{(2M+\sigma_{d})^{2}-4q^{2}}[(2M+\sigma_{d})^{2}-2q^{2}]}.
\end{eqnarray}
By using the equation (\ref{37}) we have the following equation,
\begin{equation}
c_{L}=c_{R}=\frac{6[(2M+\sigma_{d})^{2}-2q^{2}](2M+\sigma_{d})}{q^{2}}.
\end{equation}
Here we see that the universal relations help us to obtain the entropy quantities for the left and right sectors. Also in above relations we see that the left and right moving sectors of entropies could be written in the form of Cardy formula.  Such corresponding formula give us information about left and right central charges of CFT theory. So, we can say that two sides of central charge in CFT will be same. These results play important role in improving AdS/CFT and holography.

\section{Conclusion}\label{IV}
In this paper we investigated the thermodynamic properties of inner and outer horizons in Kiselev black hole solution.  We used such corresponding horizons in case of black hole surrounded by radiation and dust fields. In both cases, we calculated the radii, the surface area, the entropy, surface temperature, the Komar energy and heat capacity products for both horizons. Also, we have shown that the surface area product, the entropy product and irreducible mass product are universal quantities. The rest of product are complectly depend on the black hole mass. Thermodynamical quantities of Horava-Lifshitz black hole investigated by the Ref. \cite{HL1}. These results completely agree with the case of Horava-Lifshitz black hole which investigated by the Ref. \cite{end}. In that case, the quantum correction to the Horava-Lifshitz black hole also studied \cite{endQ}. The holography give us opportunity to employ such universal relations and obtain entropy and central charges. Finally, we have shown that the left and right central charges of CFT will be the same. For the future work it is interesting to consider entropy corrected thermodynamics \cite{S01,S02,S04,S05,S06,S07,S08} and calculate the central charge.

\end{document}